# A comparison of single-trial EEG classification and EEG-informed fMRI across three MR compatible EEG recording systems

Josef Faller, Linbi Hong, Jennifer Cummings and Paul Sajda, *Fellow, IEEE*

*Abstract—* Simultaneously recorded electroencephalography (EEG) and functional magnetic resonance imaging (fMRI) can be used to non-invasively measure the spatiotemporal dynamics of the human brain. One challenge is dealing with the artifacts that each modality introduces into the other when the two are recorded concurrently, for example the ballistocardiogram (BCG). We conducted a preliminary comparison of three different MR compatible EEG recording systems and assessed their performance in terms of single-trial classification of the EEG when simultaneously collecting fMRI. We found tradeoffs across all three systems, for example varied ease of setup and improved classification accuracy with reference electrodes (REF) but not for pulse artifact subtraction (PAS) or reference layer adaptive filtering (RLAF).

## I. INTRODUCTION

We collected data from one healthy adult who performed an auditory oddball task and compared the quality of single-trial decoding of their EEG across three difference MR compatible EEG systems. We also conducted an EEG-informed fMRI analysis (not shown due to space limitations) and compared the three systems. The three systems included (1) our custom built fMRI compatible EEG recording system ("*LIINC*", [1]), (2) the most common commercial system ("*BP*", Brain Products, Gilching, Germany) and (3) a prototype cap (g.Tec, Schiedlberg, Austria) that connects to the BP amplifier and features a layer of reference electrodes isolated from the scalp ("*GTEC-EEG/REF*") which allows for (4) the suppression of the BCG artifact based on adaptive filtering ("*GTEC-RLAF*").

## II. METHODS

We acquired EEG and whole brain fMRI (3T GE MR750; GE Healthcare, Waukesha, WI) simultaneously in separate sessions for every one of the setups. After gradient artifact (GA) removal we classified EEG brain responses to oddball and standard stimuli using logistic regression in leave-one-out cross-validation separately for every 50 ms window centered at increments of 25 ms between 0 and 1000 ms relative to stimulus onset [1]. For *GTEC-RLAF* we also performed RLAF [2]. For conventional fMRI analysis, the event-related variability and response time variability were used as basis for regressors in a general linear model. For EEG-informed fMRI analysis, single-trial variability (STV) modulated regressors based on stimulus events were added for windows where EEG-based classification was better than chance [1].

*This work was funded by the Army Research Laboratory under cooperative agreement number W911NF-10-2-0022.

J. Faller, L. Hong, J. Cummings and P. Sajda are with the Laboratory for Intelligent Imaging and Neural Computing (LIINC), New York City, NY, 10027 USA. Correspondence: josef.faller@gmail.com.

## III. RESULTS

We find comparable results for EEG-based classification between systems (see Figure 1), but RLAF (or PAS alone) did not improve classification performance. Interestingly, all systems achieved similar or higher performance during simultaneous fMRI recording, compared to a recording in an office environment. The results of traditional fMRI analysis for all three recording systems (not shown) are consistent with previous findings ([1]) and no system caused particularly strong distortions in the fMRI signal. Results from EEG-informed fMRI analysis were most consistent with previous findings for *LIINC* and in part consistent for the two *GTEC* setups [1]. No consistent STV results were found for *BP*. We also note that setup time for the three systems is significantly different with *LIINC* being about 10-15 mins while *GTEC* and *BP* > 30 mins.

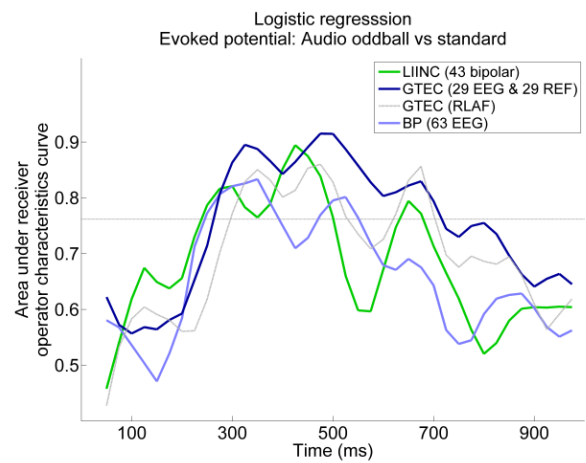

Figure 1: Stimulus-locked classification performance.

## IV. DISCUSSION AND CONCLUSION

Our findings suggest that including reference electrodes could readily improve classification performance, while artifact suppression methods like PAS or RLAF might not always lead to better results. Setup time is also a consideration and potential tradeoff. Of course data from additional subjects is necessary to derive more generalizable results.